# All-optical classification of real biomedical cell images using a diffractive neural network: a simulation study

**Norihide Sagami[1], Yueyun Weng[2], Cheng Lei[2], Ryosuke Oketani[1], Kotaro Hiramatsu[1*]**

*[1]Department of Chemistry, Faculty of Science, Kyushu University, 744 Motooka, Nishi-ku, Fukuoka 819-0395, Japan*

*[2]The Institute of Technological Sciences, Wuhan University, No. 8 Xiongtingbi Road, Xuesi Road Area, Jiangxia District, Wuhan 430200, Hubei, China*

**hiramatsu@chem.kyushu-univ.jp*

**We report an in-silico demonstration of an all-optical cell classification system using a single-layer diffractive neural network (DNN) optimized for real-world biomedical images. Implemented virtually with a spatial light modulator (SLM), the DNN was numerically trained via backpropagation to differentiate breast and lung cancer cells. The training utilized experimentally acquired phase and amplitude images from optofluidic time-stretch quantitative phase imaging. Classification was simulated by computing the optical intensities at the detection plane. The optimized DNN achieved 93.6% accuracy, approaching that of conventional convolutional neural networks. This study highlights the potential of SLM-based DNNs for ultrafast, energy-efficient biomedical image processing in practical optical computing scenarios.**

The demand for high-throughput digital signal processing is growing due to the rapid increase in digitized data not only in information technology but also across diverse natural science fields, such as bioimaging, materials informatics, pharmaceutical science, and omics research [1–4]. This surge in computational requirements has led to a dramatic increase in energy consumption, raising serious concerns regarding global energy sustainability and carbon emissions. These challenges have spurred interest in alternative computing paradigms that combine high performance with energy efficiency, such as optical computing and neuromorphic architectures [5,6].

Optical neural networks (ONNs), in particular, have gained significant attention for their potential to achieve ultrafast and energy-efficient signal processing. In recent years, a variety of ONN implementations have been proposed and demonstrated, including nanophotonic circuits, optical reservoir computing, and diffractive neural networks [7–9]. Among them, diffractive neural networks (DNNs) stand out as particularly promising architectures, enabling large-scale, high-dimensional neural computations entirely in free space. Each layer in a DNN is implemented as a diffractive optical element, with forward propagation realized passively through optical diffraction. The layer structures are digitally optimized using backpropagation simulations of wavefront propagation. As optical signals travel at the speed of light, DNNs offer inherently passive and ultrafast computation. Although DNNs currently have a narrower range of applications compared to conventional electronic neural networks, their exceptional speed and energy efficiency make them highly attractive for specialized tasks constrained by throughput and power consumption.

So far, most experimental demonstrations of DNNs for image analysis have focused on simple and standardized datasets such as MNIST handwritten digits [10–13]. In a typical DNN workflow, electronic images are converted into optical phases and/or amplitudes using spatial light modulators (SLMs) or digital micromirror devices (DMDs) and fed into the input layer of the optical neural network. While such studies convincingly validate the high-speed, energy-efficient processing capability of DNNs, they often fall short of demonstrating their practical utility in more complex and diverse real-world applications. To unlock the full potential of DNNs, it is essential to move beyond standardized benchmarks and address more complex and practical datasets that better reflect the challenges of real-world scenarios. These include tasks involving natural images, biomedical data, and multimodal inputs, where DNNs must demonstrate not only ultrafast and energy-efficient processing, but also robustness and accuracy under realistic conditions. Advancing DNNs in such demanding contexts will be key to realizing their broader adoption and impact.

In this study, we demonstrate DNN-based cell image classification in-silico. We developed a virtual single-layer DNN based on a SLM to classify breast cancer cells (BCCs) and lung cancer cells (LCCs). The network was optimized via backpropagation using a large cell image dataset, which comprised phase and amplitude images experimentally acquired by optofluidic time-stretch quantitative phase imaging (OTS-QPI) [14]. Classification was performed by measuring optical intensities at the detection plane after the input images passed through the SLM with an optimized phase pattern and amplitude masks. The optimized ONN achieved a high classification accuracy of 93.6%. These results demonstrate the potential of DNNs for high-speed, all-optical classification of cell images. Our findings verify the feasibility of DNN-based classification for real-world biomedical applications, highlighting the advantages of integrating naturally captured optical fields with diffractive computing.

The schematic of our virtual cell image classifier based on a single-layer DNN with an SLM is shown in Fig. 1. Similar to conventional DNNs, the single-layer diffractive neural network follows the principles of optical diffraction theory and machine learning systems [15,16]. In this network, forward propagation is represented by propagating the input plane's electric field according to the Rayleigh–Sommerfeld diffraction theory, and the resulting intensity distribution of the electric field on the detection plane is used as the output [17]. The obtained output is then compared with the target output to compute an error, and various parameters, including the SLM pattern, can be optimized using the backpropagation algorithm based on the defined loss function.

At the input plane, a complex electric field constructed from the phase and amplitude images of cells is provided.

$$E_{\mathrm{in}}(x,y) = A(x,y)\exp[i\phi(x,y)],$$

where $A(x,y)$ is the amplitude pattern, and $\phi(x,y)$ is the phase pattern. This field is then modulated by a trainable phase modulation pattern $\theta(x,y)$ applied by the SLM:

$$E_{\mathrm{SLM}}(x,y) = E_{\mathrm{in}}(x,y)\exp[i\theta(x,y)].$$

The modulated field propagates through free space and the propagation is simulated using the angular spectrum method:

$$E_{\mathrm{out}} = \mathcal{F}^{-1}\{\mathcal{F}[E_{\mathrm{SLM}}(x,y)] \cdot H(f_x, f_y)\},$$

where $\mathcal{F}$ denotes the two-dimensional Fourier transform and the transfer function $H(f_x, f_y)$ is defined as

$$H(f_x, f_y) = \exp\left[i2\pi z\sqrt{\lambda^{-2} - f_x^2 - f_y^2}\right],$$

where $\lambda$ is the wavelength of illumination. The resulting optical intensity $I_{\mathrm{det}}(x,y)$ at the detection plane is calculated as

$$I_{\mathrm{det}}(x,y) = |E_{\mathrm{out}}(x,y)|^2.$$

In this simulation, the propagation distance was set to $z = 15.8$ mm, corresponding to the focal length of a thin lens placed after the SLM. The wavelength of the incident light was fixed at $\lambda = 488$ nm and the spatial sampling interval of the electric field was set to $\Delta x = \Delta y = 3.45$ µm. The lens used in the model had a numerical aperture (NA) of 0.070, determining the resolution and collection angle of the diffracted light. These physical parameters were chosen to reflect realistic optical conditions and to ensure numerical stability in the simulated system.

To reduce classification sensitivity to cell orientation, a ring-shaped mask $\mathcal{M}_c(r_{\mathrm{in}}, r_{\mathrm{out}})$ is applied to the output intensity distribution $I_{\mathrm{det}}(x,y)$ at the detection plane for each class. The class-specific score is defined as the mean intensity within an annular region:

$$S_c = \frac{1}{A_c}\iint_{\mathcal{M}_c} I_{\mathrm{det}}(x,y)\, dxdy,$$

where $A_c$ is the area of the ring mask $\mathcal{M}_c(r_{\mathrm{in}}, r_{\mathrm{out}})$, defined by the inner radius $r_{\mathrm{in}}$ and outer radius $r_{\mathrm{out}}$. For binary classification between BCCs and LCCs, the final decision is based on the difference between the class scores,

$$\Delta S = S_{LCC} - S_{\mathrm{BCC}}.$$

The predicted class label is determined by the sign of $\Delta S$ : a positive value indicates LCC, and a negative value indicates a BCC. The radii $r_{\mathrm{in}}$ and $r_{\mathrm{out}}$ for each class are treated as trainable parameters and are jointly optimized with the SLM phase pattern during training.

The training dataset was prepared using the OTS-QPI system (Fig. 2a). Two types of cells were used as samples: BCCs (MCF-7 cells, obtained from the China Center for Type Culture Collection) and LCCs (A549 cells, ATCC, Manassas, VA, USA). BCCs were cultured in Dulbecco's Modified Eagle Medium (DMEM) supplemented with 9% fetal bovine serum (FBS) and 1% penicillin-streptomycin, and maintained at 37 °C with 5% $CO_2$. LCCs were cultured in Ham's F12 medium containing 9% FBS and 1% penicillin–streptomycin under the same incubation conditions (37 °C, 5% $CO_2$). Phase and amplitude images of the cells (Figs. 2b and 2c) were acquired using the OTS-QPI setup, as previously described [14]. Briefly, a broadband femtosecond laser pulse was chirped using a single-mode fiber and then split into two paths: one serving as the probe for QPI measurement, and the other as the reference. The probe beam was spectrally dispersed by a reflection grating and focused onto the sample flowing through a microfluidic device. The transmitted probe light was reflected by a second grating to cancel the spectral dispersion, then combined with the reference beam after passing through a delay line to equalize the optical path lengths. The intensity of the combined beam was measured by a high-speed photodetector. The phase and amplitude information were extracted from the temporal oscillating fringe pattern using a Hilbert transform. Because the OTS-QPI technique encodes spatial information of the cell into the time domain, the experimentally acquired temporal profiles can be converted into spatial profiles.

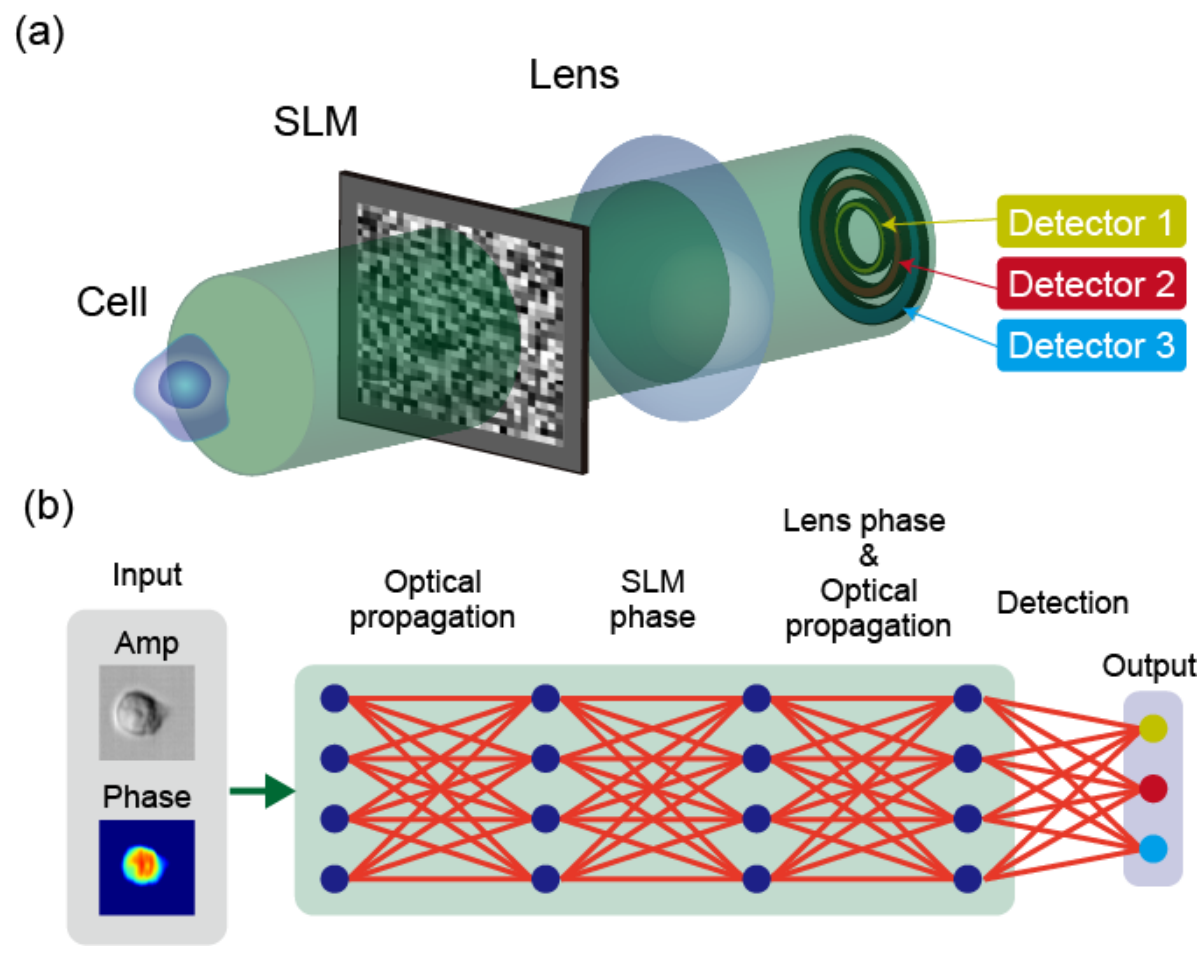

Fig. 1 Concept of optical image classifier (a) Schematic of the optical system (b) Digital model of the optical system

For each class, 2,000 images were augmented fourfold by applying horizontal and vertical flip operations. The entire dataset was split into training and validation sets in an 8:2 ratio.

We deployed a loss function based on cosine similarity to align the score difference with the ground truth label. Cosine-based loss functions have been widely adopted in classification tasks, particularly in metric learning and face recognition, for their ability to promote angular separability between feature representations [18,19]. Among various cosine-based formulations, we selected the simplest version without an explicit margin. The loss is defined as:

$$\mathcal{L} = 1 - \cos(\Delta S, y),$$

where $y \in [+1, -1]$ is the ground truth label. This formulation encourages cosine similarity between prediction and label vector, while maintaining a simple model structure and training procedure. Training was conducted using the Adam optimizer with an initial learning rate of $1.0 \times 10^3$ .The program was implemented in Python, and the neural network was developed using the PyTorch framework. To accelerate computation, an NVIDIA GeForce RTX 4090 graphics processing unit was used.

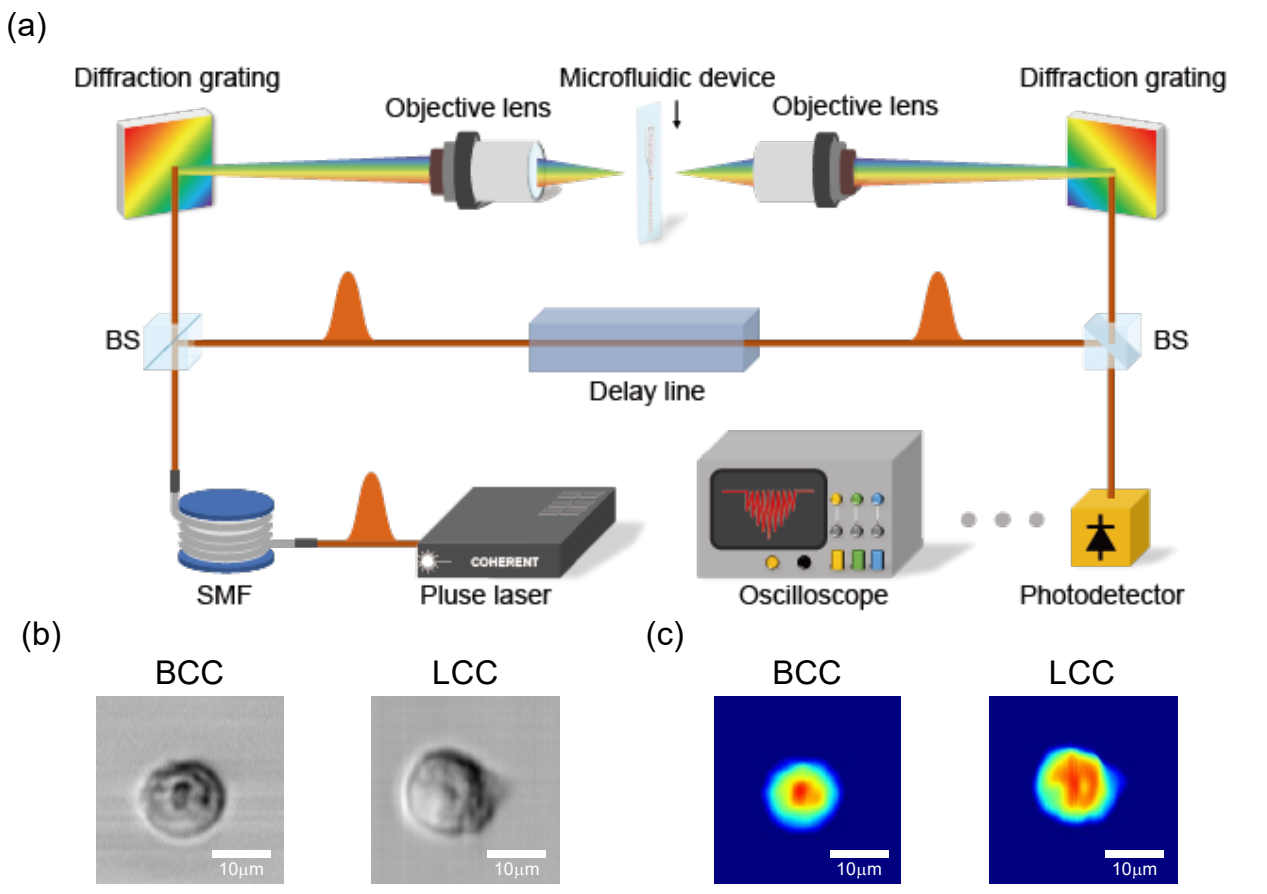

Fig. 2 Typical phase and amplitude images. (a) Setup of OTS-QPI, (b) Amplitude images, (c) Phase images

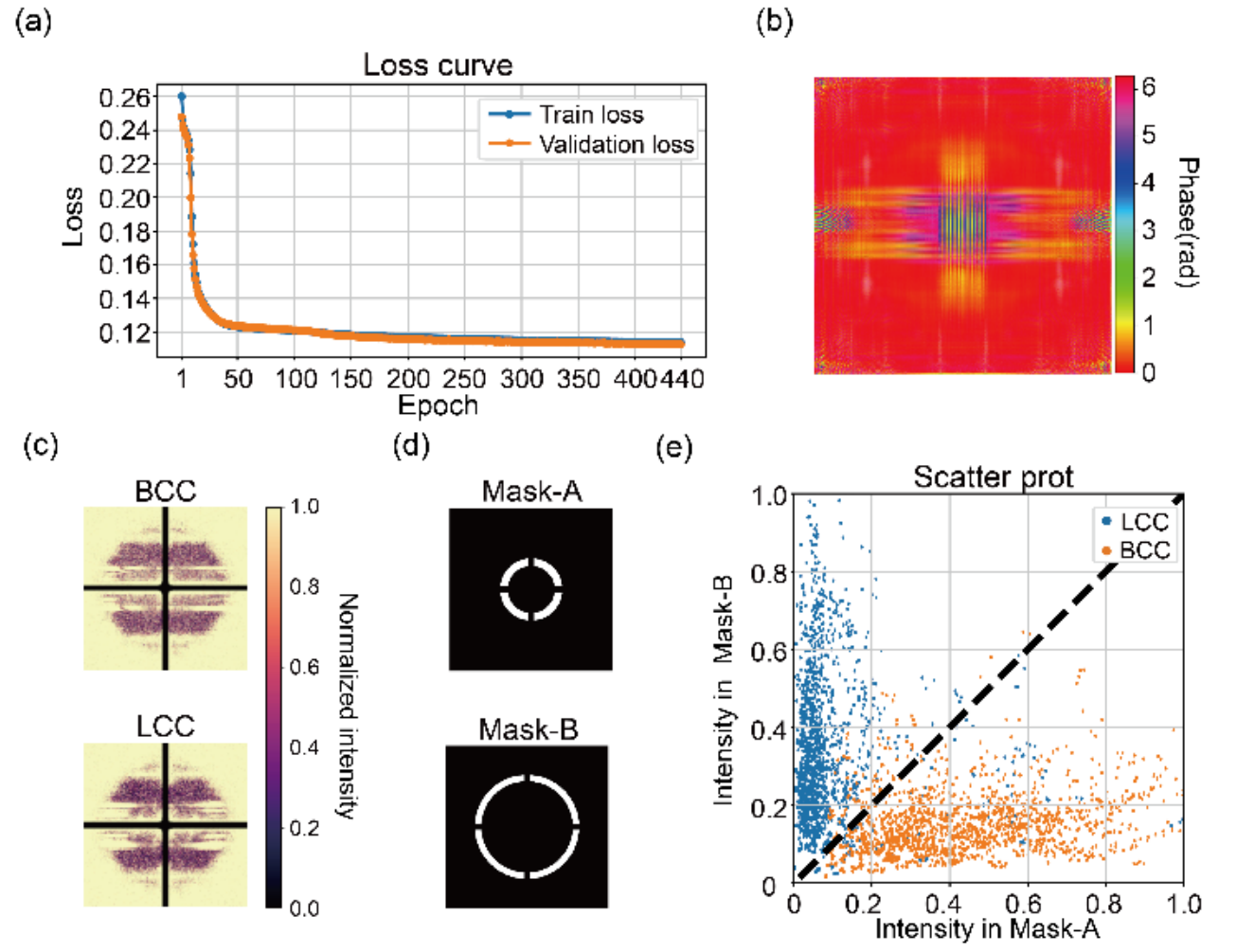

Fig. 3 Optimization of the digital model for cell classification. (a) Loss curve. (b) Optimized SLM phase pattern. (c) Ring-shaped masks for each class, optimized during training. (d) Output intensity distributions at the detection plane for BCC and LCC. A cross-shaped mask was applied to suppress central artifacts commonly observed due to optical system symmetry. (e) A scatter plot of the average intensities after applying each optimized mask to the output shown in (d). The horizontal axis represents the normalized average intensity obtained using Mask-A, while the vertical axis represents that obtained using Mask-B. The data are clipped at the 98th percentile across all samples. The black dashed line indicates the identity line.

Fig. 3(a) presents the training loss over epochs, illustrating the convergence behavior of the network. The training was terminated once the loss ceased to improve significantly, with the convergence threshold set at $1.0 \times 10^{-4}$. This indicates that the model achieved stable and sufficient optimization before reaching the maximum number of epochs. The optimized SLM phase pattern seen in Fig. 3(b) shows a complex and structured spatial pattern. This suggests that the network learned phase modulation to encode class-specific features in the optical domain. Fig. 3(d) presents representative output intensity distributions at the detection plane for BCCs and LCCs after propagation through the trained SLM. These raw outputs exhibit distinct patterns depending on the input cell type. To evaluate class separability, ring-shaped masks optimized during training (shown in Fig. 3(c)) were applied to these intensity maps. Fig. 3(e) plots the normalized average intensities obtained by applying Mask-A (horizontal axis) and Mask-B (vertical axis) to each output. For BCCs, a higher average intensity was observed in the Mask A region and lower in the Mask-B region. Conversely, LCCs showed the opposite trend, with stronger responses in the Mask-B region. These class-dependent intensities indicate that the trained DNN and ring-shaped masks separated the optical field distributions by cell type, enhancing inter-class variance in the optical feature space.

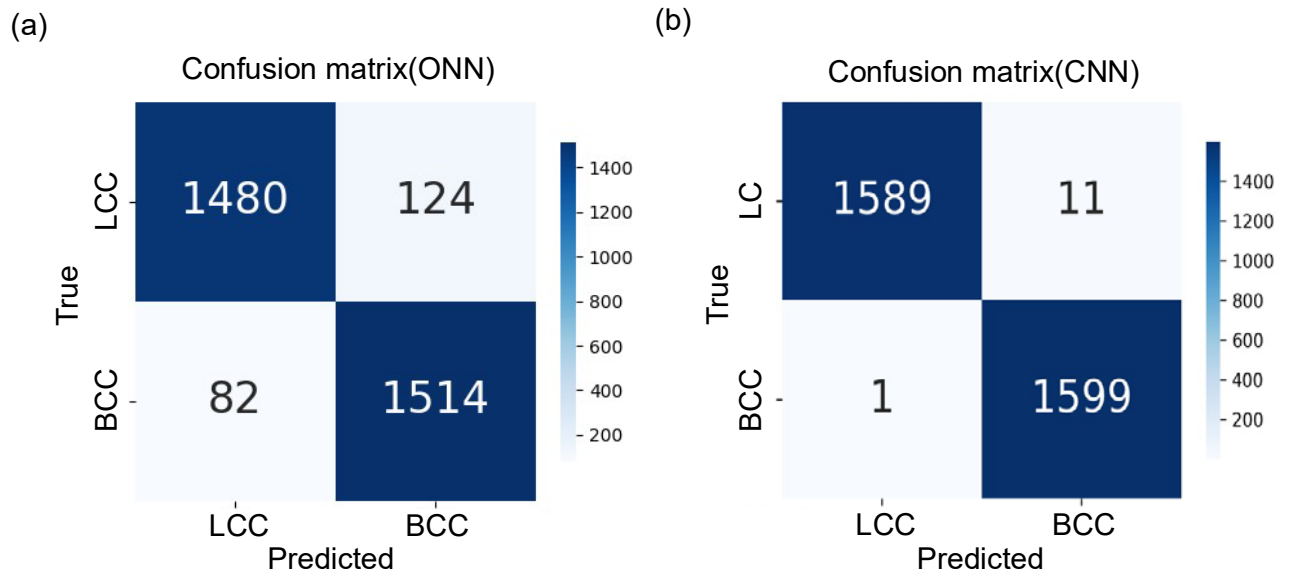

Fig. 4 Result of cell image classification (a) Confusion matrix of DNN (b) Confusion matrix of CNN

The classification performance is quantitatively evaluated in Fig. 4(a), where the confusion matrix reveals that most samples are correctly classified, with an overall accuracy of 93.6%. As a comparison, we also performed classification using a CNN. In this study, we adopted ResNet as the CNN model [20]. As a result, it achieved a classification accuracy of 99% (fig. 4(b)). This high performance is attributed to the deep architecture and rich nonlinear representation capability of ResNet. While the CNN outperforms the optical system in terms of accuracy, the all-optical classifier offers distinct advantages, including real-time and energy-efficient inference. These features make the optical approach attractive for applications where speed and energy efficiency are critical.

The results presented in this study demonstrate the feasibility of single-layer DNN for in-silico cell classification using phase and amplitude images captured by OTS-QPI. Looking ahead, several directions offer promising opportunities to enhance the capabilities and practical relevance of this approach. First, transitioning from simulation to experimental implementation is a crucial next step. Realizing a physical DNN system with SLMs and optical detection units would allow real-time, all-optical inference, validating the energy efficiency and speed benefits observed in theory. Addressing noise, aberrations, and device limitations in such physical systems will be key to robust performance in practical settings. Second, the current single-layer architecture can be extended to multilayer diffractive systems, as our simulation results suggest that increasing the number of layers improves classification accuracy. Designing compact, cascaded DNNs with optimized interlayer spacing and alignment strategies will be essential for scaling up model complexity while maintaining optical throughput and alignment tolerance. Third, the system's applicability can be broadened by integrating more diverse and clinically relevant datasets. Incorporating additional cell types, such as immune cells, circulating tumor cells, or stem cells, would enable the development of more generalizable and diagnostic tools. Moreover, multi-modal inputs, such as combined phase, fluorescence, and scattering data, could further improve classification robustness. Finally, the optimization strategy itself may benefit from recent advances in physics-informed learning, meta-learning, or reinforcement learning, which could offer better generalization and adaptive training schemes for optical hardware. Likewise, hardware-software co-design approaches could guide the joint development of optical elements and learning architectures tailored for specific biomedical tasks. By advancing in these directions, DNN-based optical computing holds significant potential for revolutionizing high-throughput, label-free cell analysis in biomedical and clinical applications.

## Acknowledgement

This work was supported by JST FOREST (JP21470594), JSPS Gran-in-Aid for Scientific Research (B) (JP23K23297, JP25K03136), Grant-in-Aid for Transformative Research Areas (JP25H01396, JP25H01393), Grant-in-Aid for Challenging Research (Pioneering) (JP25K21710), JSPS Gran-in-Aid for Scientific Research (S) (JP25H00410), Photographic Research Fund of Konica Minolta Imaging Science Foundation, and Murata Science and Education Foundation.

## References

1. G. P. Way, H. Sailem, S. Shave, R. Kasprowicz, and N. O. Carragher, SLAS Discovery **28**, 292 (2023).
2. R. Ramprasad, R. Batra, G. Pilania, A. Mannodi-Kanakkithodi, and C. Kim, npj Comput Mater **3**, 54 (2017).
3. P. Kandhare, M. Kurlekar, T. Deshpande, and A. Pawar, Medicine in Novel Technology and Devices **27**, 100375 (2025).
4. J. Lim, C. Park, M. Kim, H. Kim, J. Kim, and D.-S. Lee, Exp Mol Med **56**, 515 (2024).
5. T. Fu, J. Zhang, R. Sun, Y. Huang, W. Xu, S. Yang, Z. Zhu, and H. Chen, Light Sci Appl **13**, 263 (2024).
6. J. Robertson, P. Kirkland, J. A. Alanis, M. Hejda, J. Bueno, G. Di Caterina, and A. Hurtado, Sci Rep **12**, 4874 (2022).
7. Y. Shen, N. C. Harris, S. Skirlo, M. Prabhu, T. Baehr-Jones, M. Hochberg, X. Sun, S. Zhao, H. Larochelle, D. Englund, and M. Soljačić, Nature Photon **11**, 441 (2017).
8. D. Brunner, M. C. Soriano, C. R. Mirasso, and I. Fischer, Nat Commun **4**, 1364 (2013).
9. X. Lin, Y. Rivenson, N. T. Yardimci, M. Veli, Y. Luo, M. Jarrahi, and A. Ozcan, Science (2018).
10. H. Chen, J. Feng, M. Jiang, Y. Wang, J. Lin, J. Tan, and P. Jin, Engineering **7**, 1483 (2021).
11. D. Mengu, Y. Luo, Y. Rivenson, and A. Ozcan, IEEE Journal of Selected Topics in Quantum Electronics **26**, 1 (2020).
12. Y. Wang, A. Yu, Y. Cheng, and J. Qi, Laser & Photonics Reviews **18**, 2300903 (2024).
13. T. Yan, J. Wu, T. Zhou, H. Xie, F. Xu, J. Fan, L. Fang, X. Lin, and Q. Dai, Phys. Rev. Lett. **123**, 023901 (2019).
14. B. Guo, C. Lei, Y. Wu, H. Kobayashi, T. Ito, Y. Yalikun, S. Lee, A. Isozaki, M. Li, Y. Jiang, A. Yasumoto, D. Di Carlo, Y. Tanaka, Y. Yatomi, Y. Ozeki, and K. Goda, Methods **136**, 116 (2018).
15. Y. Rivenson, Z. Göröcs, H. Günaydin, Y. Zhang, H. Wang, and A. Ozcan, Optica, OPTICA **4**, 1437 (2017).
16. Y. Li, Y. Xue, and L. Tian, Optica, OPTICA **5**, 1181 (2018).
17. J. W. Goodman, (Roberts and Company Publishers, 2005).
18. H. Wang, Y. Wang, Z. Zhou, X. Ji, D. Gong, J. Zhou, Z. Li, and W. Liu, in (2018), pp. 5265–5274.
19. J. Deng, J. Guo, N. Xue, and S. Zafeiriou, in (2019), pp. 4690–4699.
20. K. He, X. Zhang, S. Ren, and J. Sun, in (2016), pp. 770–778.